\documentclass[12pt]{article}
\usepackage[dvips]{graphicx}
\begin{document}
\newcommand{\bea}{\begin{eqnarray}}
\newcommand{\eea}{\end{eqnarray}}
\newcommand{\bean}{\begin{eqnarray*}}
\newcommand{\eean}{\end{eqnarray*}}
\newcommand{\ba}{\begin{array}}
\newcommand{\ea}{\end{array}}
\newcommand{\be}{\begin{equation}}
\newcommand{\ee}{\end{equation}}
\newcommand{\nn}{\nonumber}

\newcommand{\bra}[1]{\langle #1|}
\newcommand{\ket}[1]{|#1\rangle}
\newcommand{\av}[1] {\langle #1\rangle}
\newcommand{\vac}[1] {\langle 0| #1|0\rangle}
\newcommand{\amp}[2]{\langle #1|#2\rangle}

\newcommand{\da}{\dagger}
\newcommand{\pa}{\partial}
\newcommand{\ga}{\gamma}
\newcommand{\ep}{\epsilon}
\newcommand{\half}{\ensuremath{\frac{1}{2}}}
\newcommand{\hh}{\hat H}
\newcommand{\ha}{\hat a}
\newcommand{\hAA}{\hat A}
\newcommand{\hN}{\hat N}
\newcommand{\hp}{\hat \psi}
\newcommand{\hphi}{\hat \phi}
\newcommand{\hpd}{\hat \psi ^\dagger}
\newcommand{\he}{\hat E}
\newcommand{\hb}{\hat b}
\newcommand{\hc}{\hat c}
\newcommand{\hjo}{\hat j _0}
\newcommand{\hr}{\hat \rho}
\newcommand{\leave}{\! \! \! \! \! / \, \,}
\newcommand{\intl}[1]{\int d\! #1 \,} 
\newcommand{\intll}[3]{\int _#1^#2 d\! #3 \,} 
\newcommand{\lm}{\lim _{y \rightarrow x}}
\newcommand{\scd}{\partial ^2 _{A_T}}

\newcommand{\fd}[1]{\frac{\delta }{\delta #1}} 
\newcommand{\pad}[1]{\frac{\partial}{\partial #1}} 
\newcommand{\refpa}[1]{(\ref{#1})} 
\newcommand{\calH}{\ensuremath{\mathcal{H}}}
\newcommand{\calD}{\ensuremath{\mathcal{D}}}
\newcommand{\calL}{\ensuremath{\mathcal{L}}}
\newcommand{\calO}{\ensuremath{\mathcal{O}}}
\newcommand{\calK}{\ensuremath{\mathcal{K}}}

\newcommand{\Tr}{\ensuremath{\mathrm{Tr}}}
\newcommand{\tr}{\ensuremath{\mathrm{tr}}}
\newcommand{\ra}{\rightarrow}
\newcommand{\lr}{\leftrightarrow}
\newcommand{\phistar}{\phi^*}
\newcommand{\etat}{\eta_T}

\newcommand{\het}{\hat E_T}
\newcommand{\hpt}{\hat \psi_T}
\newcommand{\hpdt}{\hat \psi ^\dagger_T}
\newcommand{\bfga}{\ensuremath{\mbox{\boldmath{$\gamma$}}}}
\newcommand{\bfx}{\ensuremath{\mathbf{x}}}
\newcommand{\bfy}{\ensuremath{\mathbf{y}}}
\newcommand{\bfz}{\ensuremath{\mathbf{z}}}
\newcommand{\bfp}{\ensuremath{\mathbf{p}}}
\newcommand{\bfP}{\ensuremath{\mathbf{P}}}
\newcommand{\bfq}{\ensuremath{\mathbf{q}}}
\newcommand{\bfA}{\ensuremath{\mathbf{A}}}
\newcommand{\bfB}{\ensuremath{\mathbf{B}}}
\newcommand{\bfD}{\ensuremath{\mathbf{D}}}
\newcommand{\bfE}{\ensuremath{\mathbf{E}}}
\newcommand{\bfr}{\ensuremath{\mathbf{r}}}
\newcommand{\bfOm}{\ensuremath{\mathbf{\Om}}}
\newcommand{\om}{\omega}
\newcommand{\Om}{\Omega}
\newcommand{\sgn}{\mbox{sgn}}

\newcommand{\gammat}{\tilde{\gamma}}
\newcommand{\prd}[3] {Phys.\ Rev.\ D               {#1} {(#2)} {#3}}
\newcommand{\annp}[3]{Ann.\ Phys.\ (N.Y.)          {#1} {(#2)} {#3}}
\newcommand{\pr}[3]  {Phys.\ Rev.                  {#1} {(#2)} {#3}}
\newcommand{\npb}[3] {Nucl.\ Phys.\ B              {#1} {(#2)} {#3}}
\newcommand{\prl}[3] {Phys.\ Rev.\ Lett.           {#1} {(#2)} {#3}}
\newcommand{\binom}[2] {{#1\choose #2}}

\newcommand{\hj}{\hat j}
\newcommand{\hQ}{\hat Q}
\newcommand{\hJ}{\hat J}
\newcommand{\hA}{\hat A}
\newcommand{\hH}{\hat H}
\newcommand{\de}{\delta}
\newcommand{\leri}{\leftrightarrow}
\newcommand{\llabel}[1]{\label{#1}\marginpar{#1}}

\newcommand{\figcap}[1]{\refstepcounter{figure}
        {\bf Figure \thefigure}: {\small\sl #1}}
%
\thispagestyle{empty}
\begin{center}
{\large\bf THERMAL PHASE TRANSITION IN WEAKLY\\[2ex]
	 INTERACTING,
 LARGE $N_C$ QCD}\\[5mm]
\normalsize
{\large Joakim Hallin\footnote{\noindent Email address:
tfejh@fy.chalmers.se}}\\Institute of Theoretical Physics\\
   Chalmers University of Technology and G\"oteborg  University\\
 S-412 96 G\"oteborg, Sweden \\[2ex]
{\large David Persson\footnote{\noindent Email address:
persson@theory.physics.ubc.ca}}\\
Department of  Physics and Astronomy\\
University of British Columbia\\
Vancouver, B.C. V6T 1Z1, Canada\\[4em]
\end{center}

\begin{abstract}
We consider thermal QCD in the large $N_C$ limit, mainly in 1+1 dimensions.
 The gauge coupling  is only taken 
into account to minimal order, by projection onto colour singlets.
An expression for the free energy, exact as $N_C\rightarrow \infty$, is then 
obtained. A third  order phase transition will occur. The critical temperature
depends on the ratio $N_C/L$, where $L$ is the (infinite) spatial  length.
In the high temperature limit, the free energy will approach the same
value as in the free theory, whereas we have a mesonic like  phase at low 
temperature. Expressions for the quark condensate, $
\langle \bar\Psi \Psi \rangle$, are also obtained.
\end{abstract}
\newpage
\setcounter{footnote}{0}
\setcounter{page}{1}
Considerable interest has  recently been devoted to the assumed deconfinement
phase transition in QCD at high temperature and/or density. 
Exact analytical results are incomprehensible sofar, and we have
to rely on approximate methods or computer simulations on the lattice.
It has been well known for a long time that some insight in the confinement
mechanism may be gained by considering the large $N_C$ limit, for an
$SU(N_C)$ gauge theory. This limit is particularly fruitful in
1+1 dimensions (see e.g. Refs.~\cite{tHooft:1974,CallanCG:1976,Wu:1977,
Einhorn:1976,BrowerS:1979}). QCD in 1+1 dimensions has also received 
attention recently due to the remarkable fact that different regularizations
seems to lead to different models, even within the same choice of gauge.
In his pioneering work, 't Hooft~\cite{tHooft:1974} employed a cutoff around 
the origin 
 in momentum space, that later was shown to be equivalent to a principle 
value prescription~\cite{CallanCG:1976}. Within this framework, 
no free quarks can
 propagate, but they are confined into mesons, consisting of quark anti-quark
pairs.
Wu~\cite{Wu:1977}, on the other hand,
suggested another regularization, allowing for a Wick rotation to Euclidean 
space. The bound state equation is more complicated in this case, and no
solution has yet been found. Bassetto and Nardelli with 
co-workers~\cite{BassettoG:1996,BassettoN:1997,BassettoNS:1997,BassettoCN:1997}
have compared the two regularizations, and computed for example the
expectation value of the Wilson loop. The expectation value of the Wilson loop
with Wu's regularization, was recently calculated by Staudacher and 
Krauth~\cite{StaudacherK:1997},
who showed that confinement is not enforced in this case, unlike using the 
 't Hooft regularization. It has been suggested by Chibisov and 
Zhitnitsky~\cite{ChibisovZ:1995} that the models due to different
 regularizations could show up as different phases, that makes a study of
the system at finite temperature intriguing.
In 2 dimensional QCD,  confinement is obvious as an 
infinite energy for
coloured states.\bigskip

 We shall in this letter take the  interaction into 
account only by a 
projection onto colour singlets, i.e. we neglect the coupling unless it
multiplies the infra red divergence. 
The quark contribution to the free energy in
3+1 dimensions is then easily obtained from the free energy presented here in
1+1 dimensions through the substitution, $L \int dp/(2\pi) \rightarrow
2V \int d^3 \bfp/(2\pi)^3$. The projected free energy in 3+1 dimensions
 has earlier been considered in the high
temperature limit by Skagerstam~\cite{Skagerstam:1984}.
 The full  interacting case in 2 dimensions, that is troubled by infra red
divergences,
will be more extensively considered in another 
article. Including  the interaction in 
the strict large $N_C$ limit $L/N_C  \rightarrow 0$, McLerran and 
Sen~\cite{McLerranS:1985}, and employing a different method Hansson and
Zahed~\cite{HanssonZ:1993} showed that in  the low temperature  't Hooft meson
phase, the partition function $\ln Z=
{\cal O}(L)$, due to confinement. Using a different resummation scheme for
$L/N_C \rightarrow \infty$ , McLerran and Sen~\cite{McLerranS:1985} also 
found another phase of almost free quarks, with $\ln Z={\cal O}(L N_C)$ . 
The phase transition itself was not obtainable, and could only take
place as $T  \rightarrow \infty$, i.e. $T\propto N_C$.
 By considering finite $\kappa=N_C/L$,
we are here able to monitor the phase transition exactly, and for
finite $\kappa$, it will take place at finite $T$. However, due to the
negligence of the interaction we do not have confinement in the strict sense.
Our  low temperature phase consists of quark---anti-quark pairs, showing
up for example in Bose--Einstein distribution functions, but these are not
the confined mesons of  't Hooft's, since  $\ln Z={\cal O}(L N_C)$.\bigskip

Let us now consider the unnormalized density
matrix $\hat \rho=e^{-\beta \hat H}$ of free coloured fermions having the
Hamiltonian 
\be 
\hat H=\int_p \hpd (p) h(p )\hp (p ), 
\ee 
where 
\be
h(p )=\gamma^0 (\gamma_1  p +m) , 
\ee 
and $\int _p=\int \frac{d^n p}{(2\pi )^n}$, in $n$ spatial dimensions. We shall
here mainly discuss $n=1$. In the functional representation the
expression for 
$\hat \rho$ is \cite{Hallin:1995} 
\be 
\rho (\eta_1 ^* \eta_1,\eta_2 ^* \eta_2
)=N_\beta \exp ((\eta_1^*+\eta_2^*)\Om_\beta (\eta_1+\eta_2 )+\eta_1^*
\eta_2-\eta_2^* \eta_1 ), 
\ee where 
\bea 
N_\beta &=&\exp (4N_C V\int_p \ln (\cosh
{\frac{\beta}{2}\om })), \\
\Om _\beta &=&-\tanh (\frac{\beta}{2}\om )
\frac{h}{\om} \mathbf{1}_C ,\\
\om (p )&=&\sqrt {p ^2+m^2},
\label{omdef}
\eea
and $L$ denotes the spatial length. 
In the expression for $\rho (\eta_1 ^*
\eta_1,\eta_2 ^* \eta_2 )$, summation over spinor, colour and momentum indices
is understood. Moreover $N_C$ denotes the number of colours and $\mathbf{1}_C$
 is
the unit matrix in colour space. We now wish  to calculate the partition 
function for colourless states.
This is done by the operator $\hat \pi$, 
\be 
	\hat \pi =\int du \hat \pi_u, 
\ee 
that projects onto colour singlets.
Here
$\hat\pi_u=e^{i \alpha _a \hat Q^a}$ is a general global $U(N_C)$ gauge 
transformation. The partition
 function $Z$ is then 
\be 
	Z=\tr (\hat \pi \hat \rho ).  
\ee 
With the colour charge operator
\be 
\hat Q^a=\int_p\hpd
\frac{t^a}{2} \hp ,
\ee 
the global gauge
transformation $\hat \pi_u$
is in the functional
representation: 
\be 
	\pi_u(\eta_1 ^* \eta_1,\eta_2 ^* \eta_2 )=N_u \exp
	\left[(\eta_1^*+\eta_2^*)\Om_u (\eta_1+\eta_2 )+
	\eta_1^* \eta_2-\eta_2^* \eta_1 \right] .
\ee
Here we have defined 
\be N_u=\exp \left[\frac{1}{2} \tr \ln \frac{1}{4}(u+u^{-1}+2)\right] ,
\ee 
and 
\be
\Om_u=\frac{u-1}{u+1} \mathbf{1}_s .  
\ee 
Furthermore,  $\tr=L \int_p \tr_s \tr_C$
denotes the trace over all indices, momentum , spin and colour, while
$\mathbf{1}_s$ is the unit matrix in spinor space i.e. a $2\times 2$-unit
 matrix
($4\times 4 $ in $n=3$). The
group element is $u=e^{i \alpha _a \frac{t^a}{2}}$. Multiplying $\hat \rho$ and
$\hat u$ one finds 
\bea 
\lefteqn{\bra{\eta_1^*\eta_1}\hat u \hat
\rho\ket{\eta_2^*\eta_2}= N_\beta N_u \det (\mathbf {1}+\Om_\beta \Om _u)
\times} \nn\\
&& \exp
\left\{(\eta_1^*+\eta_2^*)\frac{\Om_\beta +\Om_u}{1+\Om_\beta\Om_u} 
(\eta_1+\eta_2 )+\eta_1^*\eta_2-\eta_2^* \eta_1 \right\} .
\eea 
Finally we can then give the expression for the partition function,
\bea
Z&=&\int du\,
	 \tr (\hat \pi_u \hat \rho )=\int du \det (\mathbf{2}) N_\beta N_u 
	\det(\mathbf {1}+\Om_\beta \Om _u) \nn\\
&=& e^{-\beta E_0} \int du \exp \left\{L \int _p \tr_C \ln (1+\xi u)(1+\xi
u^{-1})\right\} ,
\eea
where $E_0=-L N_C\int_p \om$ is the vacuum energy. We have here defined 
\be 
	\label{xidef}
	\xi=e^{-\beta \om},
\ee
 and $e^{i\alpha_j}$ denotes the $N_C$ eigenvalues of the $U(N_C)$ matrix $u$. 
On functions of
eigenvalues the Haar measure $du$ takes the form
\be
\int du=\int \prod _{i=1}^{N_C} d\alpha _i \prod_{i<j}\sin ^2 
\frac{\alpha_i-\alpha_j}{2} .
\ee
This gives
\bea\label{eq:partfunc}
	Z&=& e^{-\beta E_0} \int  \prod _{i=1}^{N_C} d\alpha _i 
	\prod_{i<j}\sin ^2
\frac{\alpha_i-\alpha_j}{2} \times \nn\\
&& \exp \left[L \int _p\sum_{j=1}^{N_C}\ln (1+\xi e^{i\alpha_j})(1+\xi 
	e^{-i \alpha _j})\right].
\eea
In the large $N_C$-limit the integral over $U(N_C)$ in \refpa{eq:partfunc} 
may be
calculated by the steepest descent method. We take this limit by letting $N_C$ 
and
the spatial length  $L$ tend to infinity simultaneously keeping their quotient 
\be
\label{kappadef}
\kappa=\frac{N_C}{L},
\ee constant. We solve this by introducing the density of
eigenstates, $\varrho$, in the continuum limit~\cite{GrossW:1980,Mandal:1990},
 i.e.
\be
	\sum_{j=1}^{N_C} f(\alpha_j) \rightarrow N_C \int_0^1 dt f[\alpha(t)]=
	N_C \int_{-\alpha_c}^{\alpha_c} \varrho(\alpha) d\alpha f(\alpha).
\ee
  We find two distinct phases, depending on the  inequality
\be\label{eq:whichgap}
1-\frac{2}{\kappa}\int _p \frac{1}{e^{\beta \om}-1} \geq 0.
\ee
As long as \refpa{eq:whichgap} is satisfied we are in the zero-gap phase,
$\alpha_c=\pi$, where $\varrho$ has support over the whole circle.
When \refpa{eq:whichgap} is not satisfied we are in the one-gap phase.
In this case there is a gap in which $\varrho$ lacks support, i.e. an interval
where the eigenvalues give a vanishing contribution as 
$N_C \rightarrow \infty$.\medskip

\begin{figure}
\setlength{\unitlength}{1pt}
\begin{picture}(360.000000,144.000000)
\put(69.048493,-5.312480){\makebox(0,0)[t]{\ensuremath{0}}}
\put(113.430028,-5.312480){\makebox(0,0)[t]{\ensuremath{2}}}
\put(157.809233,-5.312480){\makebox(0,0)[t]{\ensuremath{4}}}
\put(202.190767,-5.312480){\makebox(0,0)[t]{\ensuremath{6}}}
\put(246.569972,-5.312480){\makebox(0,0)[t]{\ensuremath{8}}}
\put(290.951507,-5.312480){\makebox(0,0)[t]{\ensuremath{10}}}
\put(54.588321,3.429736){\makebox(0,0)[r]{\ensuremath{0}}}
\put(54.588321,37.214504){\makebox(0,0)[r]{\ensuremath{1}}}
\put(54.588321,71.001602){\makebox(0,0)[r]{\ensuremath{2}}}
\put(54.588321,104.788700){\makebox(0,0)[r]{\ensuremath{3}}}
\put(54.588321,138.575797){\makebox(0,0)[r]{\ensuremath{4}}}
\put(75,115){\makebox(0,0)[l]{\ensuremath{\frac{m}{T}\frac1{N_C^2}
\ln Z}}}
\put(310,-4){\makebox(0,0)[t]{\ensuremath{\frac{T}{m}}}}
\includegraphics{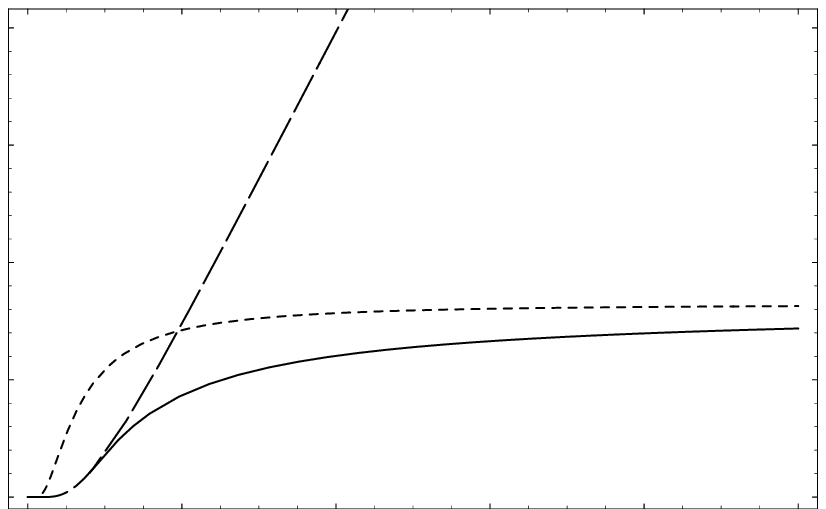}
\end{picture}\\[0.1em]

\figcap{The free energy for free quarks (dotted line), in the 0-gap phase
(dashed line), and in the 1-gap phase (solid line), for $\kappa=m$.
All in units of $N_C^2 T^2/m $.}
\label{fig-free}
\end{figure}

 In the
zero-gap phase $\ln Z$ is given by a sum over quark---antiquark pairs
\be
\ln Z=-\beta E_0 -\frac{N_C^2}{\kappa ^2}\int_p \int_q \ln \left[1-\xi (p )
\xi (q )\right]. \label{freezero}
\ee
Even though the states are mesonic-like, we do not have confinement since
$\ln Z \propto (N_C/\kappa)^2= N_C L /\kappa$, whereas in the confined
phase of `t Hooft mesons we have (cf. \cite{HanssonZ:1993}) $\ln Z \propto
L$.
In the one-gap phase we find
\bea
\ln Z&=&-\beta E_0+ N_C^2\ln \left(\sin\frac{\alpha_c}{2}\right) 
+\frac{2N_C^2}{\kappa} \int_p \ln
\frac{1}{2}(1+\xi +\sqrt{\xi ^2+2 \xi \cos \alpha _c+1})\nn\\
&&-\frac{N_C^2}{\kappa^2}
\int_p \int_q \ln \left[1-x (p ) x(q )\right]   ,
\eea
where we have defined
\be
x =\frac{1}{4 \xi \sin^2 \frac{\alpha _c}{2}} \left[1+\xi -\sqrt{\xi ^2+2
\xi \cos \alpha _c+1}\right]^2.
\ee
Furthermore, the critical angle $\alpha _c$ is determined from
\be
\label{critical}
\int _p\left(\frac{1+\xi }{\sqrt{\xi^2+2\xi \cos\alpha_c+1}}-1 \right)=\kappa
	 .
\ee
 In {\bf Figure~\ref{fig-free}}, we plot
$\ln Z/(T N_C^2)$ as a function of the temperature, for $\kappa=m$.
The critical temperature is then found as $T_c \approx 0.8  m$.
The free energy ($F=-T \ln Z$)
 in the 1-gap phase (solid line) grows as $T^2$ for large $T$,
and is approaching the free energy for free quarks, i.e. without the 
projection, (dotted line), as $T\rightarrow \infty$. Whereas the
free energy in the 0-gap phase (dashed line),
 continued above the phase transition
would grow  like $T^3$.  The high temperature behaviour in the 1-gap phase is
obtained by an expansion for $\alpha_c \ll 1$, yielding $\alpha_c \simeq
4\pi \kappa/T$. As $T\rightarrow \infty$, $\alpha_c \rightarrow 0$, i.e.
only $u=1$ gives a contribution, corresponding to no projection.\bigskip

In addition to the free energy, it is of interest to compare the entropy, and
the (renormalized) energy density in the two phases. They are given by
\bea
	s&=&\frac1L \frac{\partial}{\partial T} \left( T \ln Z \right)_L,\\
	\epsilon&=& \frac{T^2}L  \frac{\partial}{\partial T}
	 \left(  \ln Z+\beta E_0 \right)_L,
\eea
respectively. Their behaviour is similar to the that of the free energy.
In the interacting case it is well known that a nontrivial chiral
condensate appears, as shown for $m=0$ by 
Zhitnitsky~\cite{Zhitnitsky:1985,Zhitnitsky:1986}, and
in the general case by Burkardt~\cite{Burkardt:1996}. However, when the 
interaction 
only is taken into account through the projection, the vacuum is 
trivial and no chiral condensate can appear in this case.
Nevertheless, the (renormalized) 
expectation value of $\bar\psi\psi$, is an order parameter for the phase 
transition, and thus of interest. We find
\be
	\langle \bar\psi \psi\rangle = -T  \frac{\partial}{\partial m}
	 \left(  \ln Z+\beta E_0 \right)_L.
\ee
In the 0-gap phase it is straightforward to evaluate this from 
\refpa{freezero}. However, in the
1-gap phase we need to define 
\be 
	 \cot \frac{\alpha_c}{2} T \frac{\partial}{\partial m} \alpha_c 
\equiv m a,
\ee
where 
\be
	a= \left\{ \int_p \frac{\xi (1+ \xi)}{(\xi^2+2\xi 
	\cos\alpha_c+1)^{3/2}} \right\}^{-1} \times
	\left\{ \int_p \frac{\xi (1- \xi)}{(\xi^2+2\xi 
	\cos\alpha_c+1)^{3/2}}\frac1\om  \right\},
\ee
is obtained from  \refpa{critical}, using \refpa{xidef}, and \refpa{omdef}.
We then find
\be
	\frac{T}m  \frac{\partial}{\partial m} x= \frac{x}{\sqrt{\xi^2+2\xi 
	\cos\alpha_c+1}} \left[ (1+\xi )a- (1-\xi)\frac1\om\right].
\ee
This gives 
\bea
	\langle \bar\psi \psi\rangle &=& {N_C^2}m \Biggl\{ -\frac12 a
	+\frac2\kappa \int_p \frac1{1+\xi+\sqrt{\xi^2+2\xi \cos\alpha_c+1}} 
	\times \nn \\
	&& \;\; \Biggl( \left[ 1 +\frac{\xi +\cos\alpha_c}{\sqrt{\xi^2+
		2\xi \cos\alpha_c+1}}\right] \frac\xi\omega+
	\frac{2\xi \sin^2 \frac{\alpha_c}2 }{\sqrt{\xi^2+2\xi \cos\alpha_c+1}}
	\zeta \Biggr) \nn \\
	&& -\frac2{\kappa^2} \int_p \int_q \frac{x(p)x(q)}{1-
	x(p)x(q)}\; \frac{(1+\xi)a -(1-\xi)\frac1\omega}{\sqrt{\xi^2+
		2\xi \cos\alpha_c+1}} \Biggr\}.
\eea
We have been able to show that $\langle \bar\psi \psi\rangle$ is continuous 
over the phase transition, whereas higher derivatives give cumbersome 
expressions.\medskip

However, in the limit of infinite massive quarks  analytical results are
obtained. As  $m\rightarrow \infty$, we must also let $T_c  
\rightarrow \infty$, such that
their quotient
\be
	\zeta=m/T_c,
\ee
is kept fixed, in order to keep a finite particle density. 
Furthermore, we define $\tau$ according to $T=T_c \tau$.
The momentum integrals are now independent of $p$, so we write
\be
\int_p \rightarrow \Lambda.
\ee
The critical temperature is then determined from
\be
	\frac{\kappa}2= \int_p \frac1{e^{\omega/T_c}-1} \rightarrow
	\Lambda \frac1{e^{\zeta}-1}.
\ee
In the 0-gap phase we then find
\be
	\langle \bar\psi \psi\rangle=2 N_c^2 \left(\frac\Lambda\kappa 
	\right)^2 
	\frac1{e^{2\zeta/\tau}-1}.
\ee
In the 1-gap phase the critical angle is now determined from
\be
	\sqrt{1+2e^{\zeta/\tau} \cos\alpha_c +e^{2\zeta/\tau}}=
	\frac{e^{\zeta/\tau}+1}{1+\kappa/\Lambda},
\ee
that after some simplifications give
\be
	\langle \bar\psi \psi\rangle=N_c^2 \left\{ \frac12 + 
	\left(\frac\Lambda\kappa \right) \left( 2+\frac\kappa\Lambda \right)
	\frac1{e^{\zeta/\tau}+1} \right\}.
\ee
Notice here the Bose--Einstein distribution in the low temperature 0-gap phase,
as opposed to the Fermi--Dirac distribution in the 1-gap phase at high 
temperature. It is now a straightforward exercise to show that 
$\partial^2/\partial t^2 \langle \bar\psi \psi\rangle$ is discontinuous,
so that it is a third order phase transition. This is in agreement with the
study of Gattringer et al.~\cite{GattringerPS:1996} concerning static quarks
on a line. Furthermore, in the corresponding lattice model Gross and 
Witten~\cite{GrossW:1980} found a third order phase transition, so we 
conjecture that the phase transition most likely is third order also for
arbitrary mass. \bigskip

\begin{center}
ACKNOWLEDGEMENTS
\end{center}
We are grateful to Ariel Zhitnitsky for proposing the subject, as well as
stimulating discussions; and to Hans Hansson for interesting remarks, and
pointing out \cite{HanssonZ:1993}. J.~H.'s research was funded by NFR (the 
Swedish Natural Science Research Council), and D.~P.'s by STINT (the Swedish 
Foundation for Cooperation in Research and Higher Education).


\medskip

\begin{thebibliography}{10}

\bibitem{tHooft:1974}
G.~'t~Hooft,
\newblock Nucl. Phys. B {\bf B75}, 461 (1974).

\bibitem{CallanCG:1976}
C.~G. Callan, N.~Coote, and D.~Gross,
\newblock Phys. Rev. {\bf D13}, 1649 (1976).

\bibitem{Wu:1977}
T.~T. Wu,
\newblock Phys. Lett. B {\bf 71}, 142 (1977).

\bibitem{Einhorn:1976}
M.~B. Einhorn,
\newblock Phys. Rev. {\bf D14}, 3451 (1976).

\bibitem{BrowerS:1979}
R.~Brower and W.~L. Spence,
\newblock  Phys. Rev. {\bf D19}, 3024 (1979).

\bibitem{BassettoG:1996}
A.~Bassetto and L.~Griguolo,
\newblock Phys. Rev. {\bf D53}, 7385 (1996).

\bibitem{BassettoN:1997}
A.~Bassetto and G.~Nardelli,
\newblock Int. J. Mod. Phys. {\bf A12}, 1075 (1997).

\bibitem{BassettoNS:1997}
A.~Bassetto, G.~Nardelli, and A.~Shuvaev,
\newblock Nucl. Phys. {\bf B495}, 451 (1997).

\bibitem{BassettoCN:1997}
A.~Bassetto, D.~Colferai, and G.~Nardelli,
\newblock Nucl. Phys. {\bf B501}, 227 (1997).

\bibitem{StaudacherK:1997}
M.~Staudacher and W.~Krauth,
\newblock hep-th/9709101  (1997).

\bibitem{ChibisovZ:1995}
B.~Chibisov and A.~R. Zhitnitsky,
\newblock Phys. Lett. {\bf B362}, 105 (1995).

\bibitem{Skagerstam:1984}
B.~S. Skagerstam,
\newblock Z. Phys. {\bf C24}, 97 (1984).

\bibitem{McLerranS:1985}
L.~D. McLerran and A.~Sen,
\newblock Phys. Rev. {\bf D32}, 2794 (1985).

\bibitem{HanssonZ:1993}
T.~H. Hansson and I.~Zahed,
\newblock Phys. Lett.  (1993).

\bibitem{Hallin:1995}
J.~Hallin and P.~Liljenberg,
\newblock Phys.Rev. {\bf D52}, 1150 (1995).

\bibitem{GrossW:1980}
D.~J. Gross and E.~Witten,
\newblock Phys. Rev. {\bf D21}, 446 (1980).

\bibitem{Mandal:1990}
G.~Mandal,
\newblock Mod. Phys. Lett. {\bf A5}, 1147 (1990).

\bibitem{Zhitnitsky:1985}
A.~R. Zhitnitsky,
\newblock Phys. Lett. {\bf 165B}, 405 (1985).

\bibitem{Zhitnitsky:1986}
A.~R. Zhitnitsky,
\newblock Sov. J. Nucl. Phys. {\bf 44}, 139 (1986).

\bibitem{Burkardt:1996}
M.~Burkardt,
\newblock Phys. Rev. {\bf D53}, 933 (1996).

\bibitem{GattringerPS:1996}
C.~R. Gattringer, L.~D. Paniak, and G.~W. Semenoff,
\newblock Annals Phys. {\bf 256}, 74 (1997).

\end{thebibliography}
\end{document}